\documentclass[twocolumn,showpacs,prb]{revtex4}
\usepackage{graphicx}

\usepackage{amssymb}
\usepackage{amsmath}
\usepackage{dsfont}
\usepackage[latin1]{inputenc}
\usepackage{bm}% bold math
\newcommand{\vc}[1]{\bm{#1}}
\newcommand{\mat}[1]{{#1}}
\newcommand{\nn}{\nonumber}

\newcommand{\Tr}{\mathrm{Tr}}

\begin{document}

\title{Universal scaling of current fluctuations in disordered graphene}
\author{Pablo San-Jose$^{1}$, Elsa Prada$^{1}$, Dmitry S. Golubev$^{2}$}

\affiliation{ $^1$ Institut f\"{u}r Theoretische
Festk\"{o}rperphysik and DFG-Center
    for Functional Nanostructures (CFN), Universit\"{a}t Karlsruhe,
    D-76128 Karlsruhe, Germany\\
    $^2$ Forschungszentrum Karlsruhe, Institut f\"{u}r Nanotechnologie, 76021 Karlsruhe, Germany and
I.E. Tamm Department of Theoretical Physics, P.N. Lebedev Physics
Institute, 119991 Moscow, Russia}
\date{\today}

\begin{abstract}
We analyze the full transport statistics of graphene with smooth
disorder at low dopings. First we consider the case of one-dimensional (1D) disorder for
which the transmission probability distribution is given analytically
in terms of the graphene-specific mean free path. All current cumulants
are shown to scale with system parameters (doping, size, disorder
strength and correlation length) in an identical fashion for large
enough systems. In the case of 2D disorder, numerical evidence is given
for the same kind of identical scaling of all current cumulants, so
that the ratio of any two such cumulants is universal. Specific
universal values are given for the Fano factor, which is smaller than
the pseudodiffusive value of ballistic graphene ($F=1/3$) both for 1D
($F\approx 0.243$) and 2D ($F\approx 0.295$) disorders. On the other
hand, conductivity in wide samples is shown to grow without saturation
as $\sqrt{L}$ and $\log L$ with system length $L$ in the 1D and 2D
cases respectively.
\end{abstract}

\maketitle

\section{Introduction}

Stimulated by the striking results of the first experiments on graphene
flakes, \cite{Novoselov04,Novoselov05,Zhang05} interest in the problem
of transport in such system has seen extraordinary growth.
\cite{CastroNeto06,KatsnelsonReview06,Geim07} The particular subject of
disorder in graphene has been recently the center of numerous studies,
since many of the known concepts and results for transport in
disordered normal metals break down for the peculiar dispersion
relation of undoped graphene. For example, simple semiclassical
techniques fail; \cite{Aleiner06} quantum corrections can show up with
opposite sign to the conventional case (weak antilocalization) unless
valley symmetry is broken\cite{Suzuura02,McCann06}; such corrections
are conspicuously absent in the cleanest experimental samples,
\cite{Morozov06} the proposed explanation being an effective
time-reversal symmetry breaking due to curvature disorder that preserves
the valley symmetry in graphene. \cite{Morpurgo06} Indeed, the specific
symmetry properties of disorder turn out to be a crucial issue for
transport in this system. \cite{McCann06,Ostrovsky06} While
atomic-sized defects are widely thought to be of little importance for
transport at standard temperatures,
\cite{Peres06,Ziegler06,Ostrovsky06,Nomura07} smooth potentials can
have a very visible impact on the conductivity $\sigma$ of graphene.
Such electrostatic potentials can arise either from the ubiquitous
geometrical corrugation observed \cite{Meyer07} in most graphene
samples, \cite{CastroNeto06} or from ineffective screening of charges
in the environment.\cite{Hwang07,Galitski07} It has sometimes been
dubbed ``charge puddle disorder'' due to the fact that the gapless nature
of graphene makes the material respond to such a potential by forming
local particle and hole charge
accumulations.\cite{Cho07,KatsnelsonKlein06} Recent measurements
\cite{Martin07} estimate the typical size of charge puddles in the
$\xi\sim 10-30 \mathrm{nm}$ range, with a typical potential height of
$\sigma_V\sim 10 \mathrm{meV} - 100 \mathrm{meV}$.
\cite{Galitski07,CastroNeto07} This gives a typical dimensionless
disorder strength $\sigma_V\xi/\hbar v_F\approx 0.2 - 2$.

\begin{figure}
\includegraphics[width=8cm]{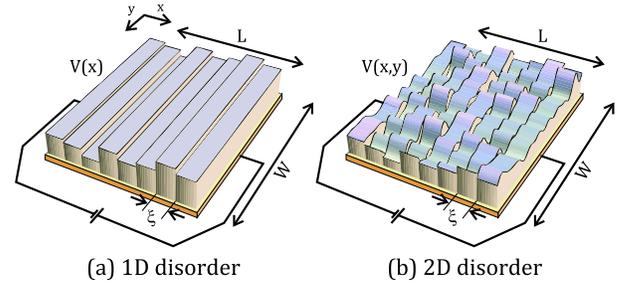}
\caption{\label{fig:disorder} (Color online) Two different disorder
realizations considered in this work for a graphene sheet (between
metallic contacts) of length $L$ and width $W$. (a) 1D disorder (no
mode coupling), (b) model for 2D charge puddle disorder.}
\end{figure}

The effect of smooth disorder on transport in graphene has been
recently analyzed theoretically by a number of authors,
\cite{Nomura07,Titov07,Rycerz07,Ostrovsky07,Bardarson07,NomuraBeta07,Cheianov07}
finding once more striking differences with respect to the well known theory
of disordered metals. \cite{BeenakkerRMP97} Due to the absence of
inter-valley scattering and to chirality conservation,
\cite{KatsnelsonKlein06} this kind of disorder has the peculiarity of
enhancing the conductivity with respect to the ballistic case, which is
at odds with classical intuition. A natural question that arises is
whether this enhancement has an upper bound and whether, as a
consequence, the conductivity of a large graphene sample exhibits a
universal value, as initially claimed, \cite{Ostrovsky07} or whether it
depends on size and system properties, in other words, whether the
scaling $\beta(\sigma)$ function in smoothly disordered graphene exists
and has fixed points or not. Some of the predictions made so far are
conflicting on this point. Using a supersymmetric nonlinear sigma
model, \cite{Ryu07} Ref. \onlinecite{Ostrovsky07} suggests the
existence of a universal minimal conductivity, whereas recent numerical
simulations find that $\sigma$ increases in a logarithmic fashion with
the system size, \cite{Bardarson07} while the beta function exists
(single parameter scaling) and always remains positive.
\cite{NomuraBeta07,Bardarson07} One must bear in mind, however, that the
conclusions of Ref. \onlinecite{Ostrovsky07} rely on a diffusive limit,
so that direct comparison to the numerical calculations might not be
straightforward. On the other hand, a recent experiment \cite{Tan07}
has shed doubts about the existence of a universal minimal conductivity
in real samples.

In this work, we analyze the effect of smooth puddle disorder on the
complete transport statistics of graphene at low dopings and, in
particular, the scaling properties of current fluctuations with system
parameters such as length or disorder strength. First, we study the case
of one-dimensional (1D) disorder. We numerically compute transport properties within the
transfer matrix formalism. We then derive and solve the single channel
Dorokhov-Mello-Pereyra-Kumar (DMPK) equation \cite{Mello88} for
graphene, valid for long samples. The DMPK equation implies that a
single parameter scaling for conductivity holds and
$\beta(\sigma)=d\log\sigma/d\log L$ exists. Both the analytical
solution and the numerical results agree with zero fitting parameters
and high accuracy. The most important difference between this result
and that of the 1D disordered metal is that in graphene, we find a
channel-dependent mean free path that scales quadratically with
transverse momentum $q$ for small dopings. The consequence of this in
transport for wider-than-long sheets is the peculiar scaling properties
of the resulting transmission probability distribution, which we
compute analytically. As a consequence of these scaling properties,
conductance and all higher current cumulants scale in the same way with
system parameters (``universal scaling''), such as system length $L$ or
disorder strength. In particular, conductivity is found to grow without
saturation as $\sqrt{L}$ with system length, as opposed to the constant
conductivity of ballistic graphene. No sign of localization is obtained
as expected. The 1D Fano factor is found to saturate, both numerically
and analytically, to $F=0.243$.

In the case of two-dimensional (2D) disorder analytical progress is difficult. We
numerically calculate the transmission probability distribution,
finding that the type of scaling features of the 1D case are also
present in two dimenions, albeit with a modified scaling law. Conductivity in
wider-than-long 2D samples is found to scale as $\sigma\propto\log L$,
like in Ref. \onlinecite{Bardarson07}, without localization, in
contrast to the initial suggestions. \cite{Ostrovsky07} All higher
cumulants once more exhibit an identical scaling law as the
conductance, leading to truly universal ratios of any pair of
cumulants, e.g., $F=0.295$.

The layout of this work is as follows. We first give an overview of the
transfer matrix method employed in this work in Sec. \ref{Transfer
matrix}. Then, we study the case of 1D disorder (Sec. \ref{1D
disorder}), and give a full analytical solution to its transport
statistics at low dopings. The same scaling is found explicitly for all
current cumulants in this case. Technical details about the 1D
calculation can be found in Appendix \ref{ap:1D}, while a derivation of
the DMPK equation in graphene can be found in Appendix \ref{ap:DMPK}. Motivated
by the 1D results, we numerically explore the case of proper 2D
disorder in Sec. \ref{2D disorder} and describe the evidence for
universal scaling of current cumulants also in two dimenions. Finally, we conclude in Sec. \ref{Conclusions}.

\section{Transfer matrix method}\label{Transfer matrix}

%Energy in this context has the meaning of average doping of the
%graphene layer.

Graphene with smooth disorder, which therefore does not couple valleys,
can be modeled by a single flavor 2D Dirac Hamiltonian
\begin{equation}
H=v_F \vc{\sigma}\cdot\vc p  + V(x,y)-\epsilon,
\end{equation}
where Pauli matrices $\vc \sigma$ act on the pseudospin space,
$\epsilon$ is the doping, $v_F\approx 10^6 m/s$ is the carrier velocity, $\vc p$ is the
momentum operator with respect to the Dirac point and $V(x,y)$ is a
disorder realization (see Fig. \ref{fig:disorder}). The two valleys and
two real spins amount to four degenerate transport channels
in this approximation. In the following, all energy scales such as
doping $\epsilon$ or potential $V$ will be given in units such that
$\hbar v_F=1$ for compactness, so that $\epsilon$ will in fact stand
for $\epsilon/(\hbar v_F)$, and so on.

In Refs. \onlinecite{Titov07} and \onlinecite{Cheianov06pn},  Titov and Cheianov and Fal'ko derived a differential equation for the transfer
matrix $\mathcal{T}_{qq'}(L)$ describing the propagation of such Dirac
fermions with transverse momenta $q=2\pi n/W$ and energy $\epsilon$
through a graphene sheet of width $W$ and length $L$ under a given
realization of disorder $V(x,y)$. In the case of large Fermi wavelength
mismatch at the contacts, it takes the form\cite{Titov07}
\begin{eqnarray}
\frac{d\mathcal{T}_{q'q}(x)}{dx}&=&\sum_{k}\mathcal{M}_{q'k}(x)\mathcal{T}_{kq}(x),\label{Teq}\\
\mathcal{M}_{q'q}(x)&=&\sigma_x\otimes q\,\delta_{q'q}+
i\sigma_z\otimes\left(\epsilon\,\delta_{q'q}-
V_{q'q}(x)\right),\label{Mq}\\
V_{q'q}(x)&=&\frac{1}{W}\int_0^W dy V(x,y) e^{-i (q'-q) y}.\label{Vq}
\end{eqnarray}
The initial condition is $\mathcal{T}(0)=1\!\!1$, and $\delta_{q,q'}$
denotes a Kronecker delta. The implicit assumptions in the above
equations are that the disorder does not couple valleys (i.e., it is
sufficiently smooth), that metallic contacts that connect graphene to
the reservoirs are appropriately modeled by infinitely doped graphene
($\epsilon\rightarrow\infty$ for $|x|>L/2$), and that the boundary
conditions for the transmission modes can be chosen periodic in the
transverse direction. The latter assumption is rigorously valid for
$W\gg L$, for which possible boundary-induced pseudospin precession and
valley mixing effects can be ignored.

The connection between the transfer and the scattering
matrix is given by
\begin{eqnarray*}
\mathcal{T}(L)=\left(\begin{array}{cc}
\hat{t}^{\dagger-1}&\hat{r}'\hat{t}'^{-1}\\
-\hat{t}'^{-1}\hat{r}& t'^{-1}\end{array}\right),
\end{eqnarray*}
where $t$ ($r$) are transmission (reflection) amplitude matrices. This
allows one to compute eigenvalues $T_n$ of the transmission matrix
$t^\dagger t$, which are conveniently recast into parameters
$\lambda_n$ through
\begin{equation}
T_n=\mbox{sech}^2 \lambda_n\, .\label{lambda}
\end{equation}
In the context of transport through random disorder $V(x,y)$, one is
interested in the probability distribution of the transfer matrix or,
more precisely, the distribution $P(\{\lambda_n\})$. Within very
general assumptions, such distribution was shown to satisfy the DMPK
equation \cite{Mello88} for disordered wires. \cite{BeenakkerRMP97} In
the next section, we show that this also applies to graphene with 1D
disorder, and we use the known solution to such 1D DMPK equation to
compute linear response transport properties.

Given the probability distribution $P(\{\lambda_n\})$ for
the transmission probabilities, a particularly useful
object is the average density distribution
$\rho(\lambda)=\sum_n\langle\delta(\lambda-\lambda_n)\rangle$
of the $\lambda_n$ above. It allows to compute any current
fluctuation average in linear response $\langle
C\rangle=\Tr C(\hat{t}^\dagger {t})$ by
\begin{equation}
\langle C\rangle=\int_0^\infty d\lambda
\,\rho(\lambda)\,C\left(\mbox{sech}^2\lambda\right).\label{cumulants}
\end{equation}
Two particularly interesting cases are the conductivity, for which
$C(T)=(4e^2/h)(L/W)T$, and the shot noise, $C(T)=(4e^3 |V|/h) T(1-T)$.
In this work, we will also consider the Fano factor, defined as the
ratio of disorder averages $F=\langle T(1-T)\rangle/\langle T\rangle$.

\section{One-dimensional disorder}\label{1D disorder}

\begin{figure}
\includegraphics[width=8cm]{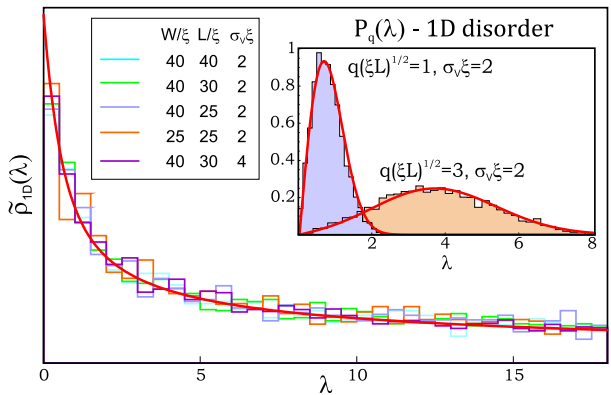}
\caption{\label{fig:rho1D} (Color online) Normalized transmission
distribution $\rho(\lambda)/\Lambda_\mathrm{1D}$ for various 1D
disordered undoped graphene systems. Good agreement is found with Eq.
(\ref{rho1D}), plotted in red. Inset: comparison of the distribution [Eq.
(\ref{P1D})] to the numerical distribution $P(\lambda_q)$ for two
different values of the transverse momentum $q$. No fitting parameters
were used in either case.}
\end{figure}

We will first use the preceding transfer matrix formalism to study the
case of one-dimensional disorder $V(x,y)=V(x)$. In this case, Eq.
(\ref{Vq}) does not mix modes, which will be labeled by a well defined
transversal momentum $q$. The corresponding $\lambda_q$ of Eq.
(\ref{lambda}) will depend on the realization of disorder but will be
otherwise mutually independent. If $V(x)$ is further modeled by
piecewise constant potential steps of width $\Delta x_i$ [see Fig.
\ref{fig:disorder} (a)], it is straightforward to compute the resulting
transfer matrix for a given $q$: $\mathcal{T}_q(L)=\prod_i
e^{\mathcal{M}_{qq}(x_i)\Delta x_i}$.

We first use this approach to numerically compute the
reduced probability distribution $P(\lambda_q)$
%$P_q(\lambda_q)\equiv\int_0^\infty \prod_{k\neq q}d\lambda_k P(\{\lambda_i\})$
for a given mode $q$. Random potential realizations are sampled with
fixed $\Delta x_i=\xi$ (representing the disorder correlation length)
and statistically independent Gaussian potentials $V(x_i)$ of variance
$\mbox{Var}[V(x_i)]=\sigma^2_V$. Each of these independent potential
steps models a 1D version of a charge puddle in graphene. \footnote{No
qualitative difference is observed in the transfer matrix using a
smoother potential.} The histogram of the computed transmissions for
each realization and mode gives the distribution $P(\lambda_q)$.

The resulting distribution $P(\lambda_q)$ for samples of length $L\gg
\xi$ (large number of puddles) is rather simple, see inset of Fig.
\ref{fig:rho1D}. For high transverse momenta $q\sqrt{\xi L}\gg 1$, it
evolves into a Gaussian centered at a large $\lambda$, while at lower
momenta, it evolves into a Rayleigh-type distribution. Such distribution
makes the limit of large number of puddles non-self-averaging, i.e., at
low momenta $\langle C(\lambda_q)\rangle\neq C(\langle
\lambda_q\rangle)$.

Analytically, one can actually prove that, in the limit of large samples
$L\gg \xi$, the distribution $P(\lambda_q)$ satisfies the single
channel DMPK equation for graphene (see Appendix \ref{ap:DMPK})
\begin{eqnarray}
l(q)\frac{\partial P}{\partial L}=
\frac{1}{4}\frac{\partial}{\partial\lambda_q}\left(\frac{\partial
P}{\partial\lambda_q} -2\coth 2\lambda_q P\right),
\end{eqnarray}
for a given mode $q$. In general, this equation involves a certain
(possibly $q$ dependent) mean free path $l$ to be determined. Its
solution is known and is given by \cite{Abrikosov81,BeenakkerRMP97}
\begin{equation}
P_{s}(\lambda)=2\sqrt{\frac{2}{\pi}}\frac{e^{-s/4}}{s^{3/2}}
\int_\lambda^\infty du
\frac{ue^{-u^2/s} \sinh 2\lambda }
{\sqrt{\cosh 2u-\cosh 2\lambda}},\label{P1D}
\end{equation}
where $s=s(q)=L/l(q)$. As a side note, this solution is,
for most practical purposes, quite indistinguishable from
the simpler Rice distribution, $P_s(\lambda)\approx
R_{s/2,\sqrt{s/2}}(\lambda)$, defined as
\begin{equation}
R_{\bar{\lambda},\nu}(\lambda)=\frac{1}{\nu^2}\exp\left(-\frac{\lambda^2+\bar\lambda^2}{2\nu^2}\right)\lambda
I_0\left(\frac{\lambda\bar\lambda}{\nu^2}\right).
\end{equation}

The DMPK equation does not describe transport through evanescent modes,
which are explicitly neglected in its derivation, see Eq. (\ref{isot}).
Transport through clean graphene at zero doping, on the other hand, is
dominated by evanescent modes. The weak disorder limit of the DMPK
solution is therefore quite different from the case of clean graphene,
which has $P(\lambda_q)\approx\delta(\lambda_q-qL)$. \cite{Tworzydlo06}
In fact, the validity of the DMPK solution requires long enough samples
so that the high $q$ evanescent modes that are unaffected by disorder
have died out, while the smaller
$|q|<\sqrt{\epsilon^2+\sigma_V^4\xi^2}$ modes are converted into
propagating modes by the effect of disorder. The ignored evanescent
modes have x-wave vector $|k_x|> \sigma_V^2\xi$, so that, more
quantitatively, this long sample condition reads $\sigma_V^2 L\xi\gg
1$. %Since the decay of such modes is exponential, the constraint is
%not very demanding, and in practice we can write it as $\sigma_V^2 L\xi
%\gtrsim 3$ to a good approximation.
The weak disorder limit of the DMPK solution corresponds, therefore, to
that in which a single charge puddle has a weak effect on the
scattering matrix, $\sigma_V\xi\ll 1$, but all of them together have a
strong effect $\sigma_V^2 L\xi\gg 1$.

 %The Gaussian and Rayleigh distributions are two
%limits of the Rice distribution, which frequently appears
%in the context of random processes in two dimensions. It
%has the form
%\begin{equation}
%R_{\bar{\lambda},\nu}(\lambda)=\frac{1}{\nu^2}\exp\left(-\frac{\lambda^2+\bar\lambda^2}{2\nu^2}\right)\lambda
%I_0\left(\frac{\lambda\bar\lambda}{\nu^2}\right),
%\end{equation}
%where $I_0$ is the modified Bessel function of the first
%kind with order zero. We therefore propose the ansatz
%\begin{equation}
%P_q(\lambda_q)\approx R_{\bar{\lambda},\nu}(\lambda_q),\label{Rice}
%\end{equation}
%given a proper choice for
%$\bar\lambda(q,\epsilon,\sigma_V)$ and
%$\nu(q,\epsilon,\sigma_V)$.
One way to find the correct form of $s(q)$ in Eq. (\ref{P1D}) is to
directly derive the DMPK equation from Eq. (\ref{Teq}). This is done in
Appendix \ref{ap:DMPK} in the case of weak disorder and small doping,
$\sigma_V\xi,\epsilon\xi\ll 1$. A more powerful way, also applicable to
strong disorder, $\sigma_V\xi\gtrsim 1$, is to obtain an exact result
for some expectation values of a function of $\lambda_q$, such as
$\langle \cosh 2\lambda_q\rangle$. It is possible to compute this
average analytically from Eq. (\ref{Teq}) in the case of 1D piecewise
disorder with a large number of charge puddles, see Appendix
\ref{ap:1D}. By comparing it to the same average obtained from the
distribution  [Eq. (\ref{P1D})], one arrives at the relation
\begin{eqnarray}
s(q)&=&\frac{1}{2}q^2 L \xi\gamma,\label{gamma}
\end{eqnarray}
for $L\gg \xi$ and $\epsilon\xi\ll 1$, where
\begin{eqnarray}
\gamma=
\frac{2(\sigma_V\xi)^2}{(\epsilon\xi)^2+(\sigma_V\xi)^4},
\label{gweak}
\end{eqnarray}
if $\sigma_V\xi\ll 1$, $\sigma_V^2\xi L\gg 1$ (we call this `weak
disorder'), and
\begin{eqnarray}
\gamma=
\frac{(\sigma_V\xi)^2(2\sqrt{2\pi}\sigma_V\xi+\pi-2)-(\epsilon\xi)^2(\pi+\sqrt{2\pi}\sigma_V\xi)}{2(\sigma_V\xi)^4},
\label{gstrong}
\end{eqnarray}
if $\sigma_V\xi\gtrsim 2$ (strong disorder). All the dependence on
doping and disorder details is contained in $\gamma$. The derivation of
the above expressions for $\gamma$ is valid for piecewise gaussian
disorder to second order in the average doping $\epsilon$ and $L\gg
\xi$. Note that an identical result is non-trivially obtained if a
different average such as $\langle \cosh^2 2\lambda_q\rangle$ is used
for the calculation, which indicates that Eq. (\ref{P1D}) is not merely
an approximation, but is indeed exact for small dopings $\epsilon\xi\ll
1$, both in the strong and weak disorder limits. We conjecture that any
other model of 1D smooth-disorder would satisfy, for large enough
samples and low dopings, the above power-law equation for $s(q)\propto
q^2$ with a certain $\gamma$ (independent of $q$ and $L$), regardless
of the details of disorder.

Equation (\ref{P1D}) is valid for any ratio $W/L$. In the case of
wider-than-long graphene sheets, however, the average eigenvalue
density function $\rho_\mathrm{1D}(\lambda)=\sum_q P_q(\lambda)$ takes
on a remarkably simple form
\begin{eqnarray}
\rho_\mathrm{1D}(\lambda)&=&\sqrt{\frac{2 W^2}{\pi L\xi
\gamma}} \tilde{\rho}_\mathrm{1D}(\lambda),\label{rho1D}\\
 \tilde{\rho}_\mathrm{1D}(\lambda)&=&\frac{\rho_\mathrm{1D}(\lambda)}{\rho_\mathrm{1D}(0)}=\frac{\sqrt{2}}{\pi}\int_\lambda^\infty
 \frac{K_1(u)\sinh 2\lambda}{\sqrt{\cosh 2u-\cosh
 2\lambda}}du,\nn
\end{eqnarray}
valid in practice for $W\gtrsim L$. Here, $K_1(u)$ is the first modified
Bessel function of the second kind. Note that for most practical
purposes, the simpler function $\tilde{\rho}_\mathrm{1D}(\lambda)\approx
e^{-\lambda}I_0(\lambda)$, where $I_0$ is the modified Bessel function
of the first kind, can be used to within excellent accuracy. Implicit
in the derivation of Eq. (\ref{rho1D}) is that all modes
$-\infty<q<\infty$ are properly described by Eq. (\ref{P1D}), which as
discussed above is valid if $\sigma_V^2\xi L \gg 1$.

In Fig. \ref{fig:rho1D}, we compare the
numerical $\rho_\mathrm{1D}(\lambda)$ as obtained by
disorder sampling in 1D to the above analytical expression
[Eq. (\ref{rho1D})] without any fitting parameters, finding a
very good agreement, even when $W$ is not much bigger than
$L$.

\begin{figure}
\includegraphics[width=8cm]{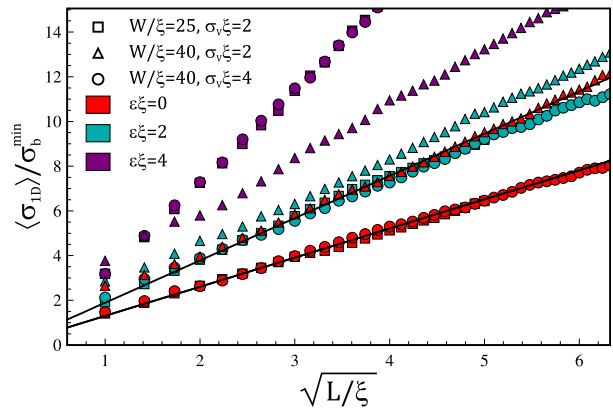}
\caption{\label{fig:sigma1D} (Color online) Numerical results for the
average conductivity vs. system length $L/\xi$ through a graphene sheet
with 1D disorder. (Note that data with $L>W$ are not shown.) The results
are normalized to the ballistic minimal conductivity
$\sigma_\mathrm{b}^\mathrm{min}=4e^2/\pi h$. A
$\sigma\propto\sqrt{L/\xi}$ scaling is obtained already from
$L\approx\xi$. At low dopings, Eq. (\ref{sigma1D}) -black lines- agrees
with the numerical results.}
\end{figure}

\begin{figure}
\includegraphics[width=8cm]{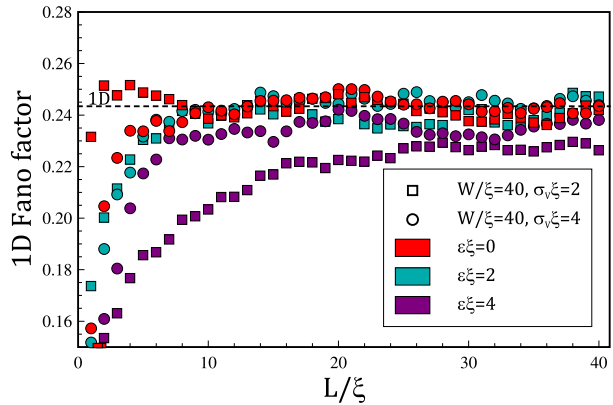}
 \caption{\label{fig:Fano1D} (Color online) Fano factor vs sheet length
$L$ for various system parameters and 1D disorder. The saturation value
is universal at low dopings (in red) and lies at the value
$F=0.243$ derived from Eq. (\ref{rho1D}).}
\end{figure}

One of the main features of $\rho_\mathrm{1D}$ for graphene with smooth
disorder [Eq. (\ref{rho1D})] is the absence of a localization length.
In other words, conductivity defined as $\sigma\equiv G \frac{L}{W}$
grows indefinitely with system size, scaling as
\begin{equation}
\sigma=0.49\sqrt{\frac{1}{\gamma}}\sqrt{\frac{L}{\xi}}\label{sigma1D}
\end{equation}
for large $L/\xi$. A similar scaling in one dimension was found in the case of
white noise disorder within an approximate self-averaging
assumption.\cite{Titov07} This is in contrast to the constant $\sigma$
of ballistic graphene \cite{Tworzydlo06} and to the $\sigma\sim\log L$
scaling behavior observed for 2D disorder by Bardarson \textit{et al.},
\cite{Bardarson07} as we will discuss in the next section. The
$L^{1/2}$ scaling of conductivity is not surprising, however, in view
of the property $l(q)\propto q^{-2}$. Indeed, within the window of
propagating $q$ modes, those with sizable transmission (ballistic
modes) satisfy $l(q)\sim 1/\xi q^2\gtrsim L$. Therefore, there will be
approximately $W/\sqrt{\xi L}$ such ballistic modes (those with smaller
q). These modes will dominate transport, so that one expects the same
$\sqrt{1/L}$ scaling for the conductance and hence a conductivity
$\sigma\propto \sqrt{L}$.

Another important feature of Eq. (\ref{rho1D}) is that all the details
of the system (disorder strength, size, and doping) enter as a
$\lambda$-independent prefactor to the density
$\tilde\rho_\mathrm{1D}(\lambda)$ and do not affect the shape of the
density profile. This directly implies that, in the parameter regime of
validity for Eq. (\ref{rho1D}), the ratio of any pair of current
cumulants is \emph{universal} (independent of system size, mean free
path, and disorder strength) since all current cumulants will scale
with these parameters in the same way as the conductivity. In
particular, the Fano factor in the presence of 1D puddle disorder close
to the Dirac point becomes $F_\mathrm{1D}\approx 0.243$, below the
pseudodiffusive prediction $F=1/3$ for ballistic
graphene.\cite{Tworzydlo06} The numerical results for the conductivity
and Fano factor in 1D are shown in Figs. \ref{fig:sigma1D} and
\ref{fig:Fano1D}, respectively, as a function of system length $L/\xi$.
The numerical Fano factor is indeed seen to saturate close to $F=0.243$
independent of system parameters. The average conductivity also
scales as $\sqrt{L/\xi}$ at low dopings as expected. Interestingly,
this scaling persists also at higher dopings for which the derivation
of $s(q)$ [Eq. (\ref{gamma})] ceases to be valid.

\section{Two-dimensional disorder}\label{2D disorder}

In a realistic model of disordered graphene with $W>L$,
mode mixing becomes important, so the preceding discussion
of 1D disorder need not apply. %Indeed, various results were
%previously found for the dependence $\sigma(L)$ in 2D that
%are markedly weaker than $\sigma(L)\sim\sqrt{L}$.
Indeed, as mentioned in the Introduction, some authors
claimed \cite{Ostrovsky07} that large graphene sheets would,
in fact, reach a universal value of conductivity in the
presence of smooth disorder (no intervalley scattering),
while others \cite{Bardarson07} numerically found a
nonuniversal conductivity scaling as $ \sigma\propto \log
L$ for deltalike and Gaussian-correlated 2D disorder
potentials. As we will show in this section, our results
support the latter nonuniversal conductivity in puddle
disordered graphene. In addition, our results suggest the
existence of universal ratios of current cumulants in the
limit $W\gtrsim L\gg\xi$, just as in the 1D case.

In the presence of mode mixing due to two-dimensional
disorder, the transfer matrix technique can still be used
for numerical simulations, although it becomes more
computationally intensive.
%Mode mixing induces effective
%interactions between the different transmission
%eigenvalues, so that the evolution of different $\lambda$'s
%is no longer independent for a given realization
%(transmission eigenvalues will now \emph{interact}, in the
%language of random matrix theory\cite{BeenakkerRMP97}).
%Here, as in Ref. \onlinecite{Bardarson07}, we use this
%method while sampling 2D disorder realizations.
Our goal once more is to compute the density
$\rho_\mathrm{2D}(\lambda)=\sum_n P_n(\lambda)$ in the presence of
charge puddle disorder of typical size $\xi$ in both $x$ and $y$
directions, which we model by a random potential profile $V(x,y)=\sum_q
a_q(x) e^{i q y}$. Harmonics $a_q(x)$ are approximated by piecewise
constant functions in $x$ at intervals of size $\xi$. For a given $x$,
$a_q(x)$ is assumed to be Gaussian distributed with variance
$\mathrm{Var}[a_q(x)]=\sigma_V^2$ for all $q<2\pi/\xi$. Higher
harmonics are suppressed, as expected by the smoothness of charge
puddles of typical size $\xi$. The resulting potential, depicted in
Fig. \ref{fig:disorder}(b), is convenient for numerical computations
[$V_{q'q}(x)=a_{q'-q}(x)$)] but can still be considered realistic for
charge puddle disorder. For practical calculations, a high-momentum
cutoff must be introduced; it is chosen high enough so that the result
for $\rho_\mathrm{2D}(\lambda)$ is cutoff independent. The computation
of the transmissions is performed by composition of scattering
matrices, rather than by multiplication of transfer matrices, since the
latter method is unstable when many modes (higher momenta) have small
transmissions, as is the case here.

\begin{figure}
\includegraphics[width=8cm]{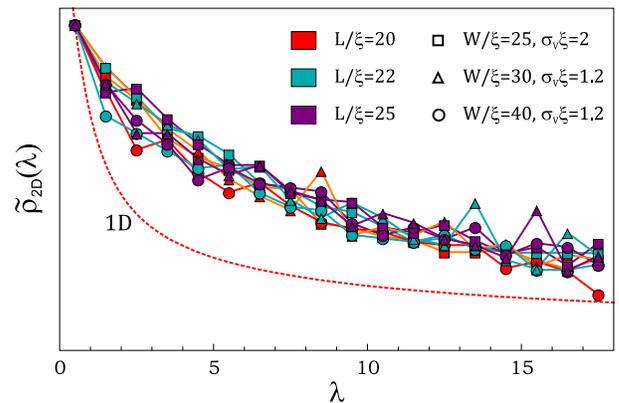}
 \caption{\label{fig:rho2D} (Color online) Normalized
 transmission density distribution for 2D disorder and zero doping. For
 large enough systems, all the curves coincide (within the
 noise), suggesting a universal scaling for current
 cumulants just as for 1D disorder (red dotted line).}
\end{figure}

As can be seen in Fig. \ref{fig:rho2D}, the density
$\rho_\mathrm{2D}(\lambda)$ clearly deviates from the 1D solution
(red dotted line). Strikingly, however, the shape of
$\rho_\mathrm{2D}(\lambda)$, up to a global rescaling factor
$\Lambda_\mathrm{2D}$ (see below), is clearly independent of the
sample length, width, or scattering strength as long as $L/\xi\gg 1$
(complete mode mixing) and $W\gtrsim L$ (continuum of transport
momenta).
When such conditions are satisfied, the disordered system acquires
some universal features. All the dependence of
$\rho_\mathrm{2D}(\lambda)$ on system parameters $W,L,\xi$,
$\sigma_V$, and $\epsilon$ appears to enter as a
$\lambda$-independent prefactor,
\begin{equation}
\rho_\mathrm{2D}(\lambda)\approx\Lambda_\mathrm{2D}(L,W,\xi,\sigma_V,\epsilon)\tilde{\rho}_\mathrm{2D}(\lambda),\label{rho2D}
\end{equation}
for some pure function of $\lambda$,
$\tilde{\rho}_\mathrm{2D}(\lambda)$, just as in the 1D case Eq.
(\ref{rho1D}). It is interesting to compare this pure function, plotted
in Fig. \ref{fig:rho2D}, to the Dorokhov result
\cite{Dorokhov84,Nazarov94} for diffusive metals which is
$\lambda$ independent for large samples. Ballistic graphene (zero
disorder) also has a $\lambda$-independent $\rho(\lambda)$.

\begin{figure}
\includegraphics[width=8cm]{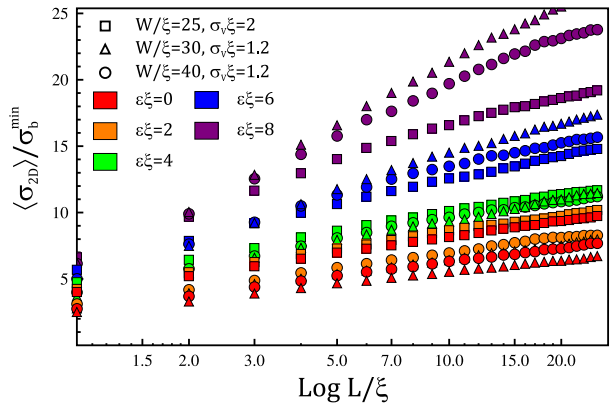}
 \caption{\label{fig:sigma2D} (Color online) Conductivity
 in 2D disordered graphene for various average dopings
 $\epsilon$ and system parameters. A clear $\log L$
 scaling is observed up to $L\sim W$, where the transition to
 a quasi-1D ribbon regime begins.}
\end{figure}
\begin{figure}
\includegraphics[width=8cm]{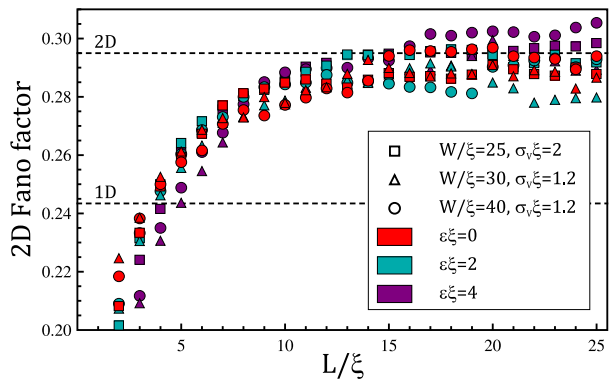}
 \caption{\label{fig:fano2D} (Color online) As for 1D disorder, the Fano factor saturates to a system independent value, this time around $F=0.295$ for large systems sizes.}
\end{figure}

Conductance and higher current cumulants are obtained from Eq.
(\ref{cumulants}). In contrast to the 1D case, we see in Fig.
\ref{fig:sigma2D} that the minimal conductivity $GL/W$ is proportional
to $\log L/\xi$, in agreement with Ref. \onlinecite{Bardarson07}. The
observed logarithmic dependence of the conductance on the sample size
$L$ might be attributed to the weak antilocalization. According to Ref.
\onlinecite{McCann06} (see also Ref. \onlinecite{Cheianov07}), the weak
antilocalization correction to the conductance of a 2D disordered
graphene in the diffusive regime $L\gg l$, with no intervalley
scattering, no magnetic impurities, and no interaction between
electrons, should read
\begin{eqnarray}
\delta\sigma \approx \frac{2e^2}{\pi^2\hbar}\ln\frac{L}{l}\equiv
\sigma_b^{\min}\ln\frac{L}{l}.
\end{eqnarray}
Our low energy numerical data appear to be in reasonable agreement with
this expression. At $\epsilon\xi < 4$, we find $ d\sigma / d\ln L =1.3 -
1.5\; \sigma_b^{\min}$  (see Fig. 6). Unfortunately, we do not know the
value of the mean free path $l$ and therefore cannot independently
verify if our samples are in the diffusive regime. Most probably, at
low energies, we get $l\lesssim L$ and the diffusive model applies,
while at higher energies, we have $l\sim L$ which leads to a stronger
dependence of the conductance on $L$. Such an interpretation of the
results plotted in Fig. \ref{fig:sigma2D} suggests that the mean free
path $l(\epsilon)$ should grow with energy as it did in case of 1D
disorder.

From the logarithmic dependence of the conductivity, it immediately
follows that $\Lambda_\mathrm{2D}(L,W,\xi,\sigma_V,\epsilon)\propto
\frac{\log (L/\xi)}{L}$. This implies that shot noise and higher
current cumulants should all scale with $\frac{\log (L/\xi)}{L}$. This
behavior is indeed observed numerically (not shown explicitly),
confirming Eq. (\ref{rho2D}).

Just as in the 1D case, the form of Eq. (\ref{rho2D}) implies once more
that the ratio of any pair of current cumulants will be universal if
$L/\xi\gg 1$ and $W\gtrsim L$. The Fano factor, in particular, acquires a
universal value $F_\mathrm{2D}\approx 0.295$, as can be seen in Fig.
\ref{fig:fano2D}. Note that $1/3>F_\mathrm{2D}>F_\mathrm{1D}$.

\section{Conclusions}\label{Conclusions}

In conclusion, we have characterized the full transport statistics of
disorder graphene in the absence of intervalley scattering. We have
computed numerically the transmission probability distribution both for
1D and 2D disorders. In both the 1D and 2D cases we have found
non-self-averaging statistics and no localization length. All current
cumulants are seen to grow monotonously with system length without any
saturation, which suggests that experimental results for the minimal
conductivity should be limited by contact resistance.

In the 1D case, we have shown that the solution for long samples
satisfies a single channel DMPK equation, for which we give the exact
analytical solution and the expression for the mean free path at low
dopings. Such mean free path scales as $l\propto q^{-2}$ with the
transverse momentum, which has important consequences for the scaling
properties of current cumulants. Such scaling properties, observed
numerically also in 2D,  suggest that, for long and wide enough
samples, all cumulants scale in an identical fashion with system size,
leading to universal ratios of any two cumulants. Although the same
phenomenon is obtained in the theory of diffusive
metals,\cite{Nazarov94} the resulting transport statistics is less
trivial in the case of graphene. The Fano factor in 2D disordered
graphene $F=0.295$, for example, is below the value in diffusive metals
and clean graphene $F=1/3$.

\acknowledgements

We would like to thank  P. Ostrovsky, I. V. Gornyi, and M.
Titov for valuable discussions and F. Guinea for his guidance and for
encouraging this study. E.P would also like to acknowledge useful discussions with V. I. Fal'ko. E.P. has benefited from the financial support of the European Community under the Marie Curie Research Training
Networks, ESR program.

\appendix

\section{Derivation of the one-dimensional Dorokhov-Mello-Pereyra-Kumar equation for graphene\label{ap:DMPK}}

Following Titov, \cite{Titov07} we parametrize the $T-$matrix as
follows:
\begin{eqnarray}
{\cal T}=\left(
\begin{array}{cc}
e^{i\varphi_1}\cosh \lambda_q & e^{-i\varphi_2}\sinh\lambda_q \\
e^{i\varphi_2}\sinh\lambda_q &
e^{-i\varphi_1}\cosh\lambda_q
\end{array}
\right).
\end{eqnarray}
From Eq. (\ref{Teq}), we derive the following equations \cite{Titov07}
\begin{eqnarray}
\frac{d \lambda_q}{d x}&=& q \cos 2\theta_q,\label{eq1a}\\
\frac{d\theta_q}{dx}&=&\epsilon-V(x)-q\sin 2\theta_q \;\coth
2\lambda_q, \label{eq1b}
\end{eqnarray}
where $\theta_q=(\varphi_1-\varphi_2)/2$. The initial condition for
Eqs. (\ref{eq1a}) and (\ref{eq1b}) reads $\lambda_q(0)=0$ and
$\theta_q(0)=0$. Our aim is to derive an equation for the averaged
distribution function $P(\lambda_q,\theta_q)$. We proceed along the
standard route and begin with the equation for the non-averaged
distribution function $\tilde
P(\lambda_q,\theta_q)=\delta[\lambda_q-\lambda_q(x)]\delta[\theta_q-\theta_q(x)]$,
where $\lambda_q(x),\theta_q(x)$ are the solution of Eqs. (\ref{eq1a}) and
(\ref{eq1b}). This equation reads
\begin{eqnarray}
\frac{\partial \tilde P}{\partial x}&=&
-\frac{\partial}{\partial\theta_q}\big(
(\epsilon-V(x)-q\sin 2\theta_q \;\coth 2\lambda_q)\tilde P
\big) \nonumber\\ &&
-\frac{\partial}{\partial\lambda_q}\big(q\cos
2\theta_q\;\tilde P\big). \label{eq2}
\end{eqnarray}
Now we assume that $q\xi,\epsilon\xi,\sigma_V\xi\ll 1$. In this case,
the fluctuating potential can be treated as $\delta$ correlated, i.e.,
we consider
\begin{eqnarray}
\langle V(x_1)V(x_2)\rangle=\sigma_V^2\xi
\delta(x_1-x_2)\label{deltacorrelated-potential}.
\end{eqnarray}
The averaging over $V(x)$ in Eq. (\ref{eq2}) becomes very
simple and we arrive at the usual Fokker-Plank
equation for $P=\langle\tilde P\rangle$:
\begin{eqnarray}
 \frac{\partial P}{\partial x}&=& -\epsilon\frac{\partial P}{\partial\theta_q}
+\frac{\sigma_V^2\xi}{2} \frac{\partial^2 P}{\partial
\theta_q^2} -q \cos 2\theta_q\frac{\partial
P}{\partial\lambda_q} \nonumber\\ && +\, q \coth
2\lambda_q\left( 2\cos 2\theta_q P +\sin 2\theta_q
\frac{\partial P}{\partial\theta_q}\right). \label{FP}
\end{eqnarray}

Next, we assume
\begin{eqnarray}
q^2\lesssim \epsilon^2+\sigma_V^4\xi^2. \label{isot}
\end{eqnarray}
Then, the isotropization of $P(\lambda_q,\theta_q)$ over the angle
$\theta_q$ happens fast, so that derivatives with respect to $\theta_q$
in Eq. (\ref{FP}) are small. We split the distribution
$P(\lambda_q,\theta_q)$ into the sum of an isotropic part and small
anisotropic correction,
\begin{eqnarray}
P(\lambda_q,\theta_q)=P(\lambda_q)+\alpha(\lambda_q)\cos
2\theta_q+\beta(\lambda_q)\sin 2\theta_q.
\end{eqnarray}
We derive three coupled equations for $P$, $\alpha$ and $\beta$. To
this end, we first average Eq. (\ref{FP}) over the angle
$\theta_q$, then multiply it with $\cos 2\theta_q$ and average the
result, and, finally, we multiply Eq. (\ref{FP}) with $\sin 2\theta_q$
and average over $\theta_q$ again. We then get
\begin{eqnarray}
\frac{\partial P}{\partial
x}&=&-\frac{q}{2}\frac{\partial\alpha}{\partial\lambda_q},
\nonumber\\
\frac{1}{2}\frac{\partial \alpha}{\partial x}&=& -\epsilon\beta -
\sigma_V^2\xi\alpha -\frac{q}{2}\frac{\partial P}{\partial\lambda_q}
+q\coth 2\lambda_q \, P,
\nonumber\\
\frac{1}{2}\frac{\partial \beta}{\partial x}&=& \epsilon\alpha -
\sigma_V^2\xi\beta.
\end{eqnarray}
Under the condition [Eq. (\ref{isot})] we can set $\partial\alpha/\partial
x=\partial\beta/\partial x=0$. Then $\alpha$ and $\beta$ are easily
excluded and we arrive at the one channel DMPK equation
\begin{eqnarray}
l(q)\frac{\partial P}{\partial x}=
\frac{1}{4}\frac{\partial}{\partial\lambda_q}\left(\frac{\partial
P}{\partial\lambda_q} -2\coth 2\lambda_q P\right),
\label{DMPK}
\end{eqnarray}
with the effective mean free path
\begin{eqnarray}
l=\frac{\epsilon^2+\sigma_V^4\xi^2}{\sigma_V^2\xi q^2}. \label{l}
\end{eqnarray}
Let us briefly discuss this result. First of all Eq. (A11) predicts an
infinite mean free path for an electron moving perpendicular to the
potential barriers ($q=0$), i.e. no back-scattering occurs in this
case. This is a manifestation of Klein paradox. At finite values of $q$
the momentum of an incident electron is no longer perpendicular to the
surface of the barrier, which makes the back-scattering possible. As a
result the mean free path (A11) becomes finite. Such behavior of the
mean free path is encoded in the mathematical structure of Eqs.
(\ref{Teq}) and (\ref{Mq}). Namely, it is related to the fact that, in
the case of 1D disorder, the off-diagonal matrix elements of ${\cal
M}_{q'q}$, responsible for back-scattering are proportional to $q$. One
can actually derive the mean free path (A11) directly from Eqs.
(\ref{Teq})-(\ref{Vq}) in a simple way. To this end one should assume
$q$ to be sufficiently small and treat the off-diagonal elements of the
matrix ${\cal M}_{q'q}$ perturbatively. Since Eq. (\ref{Teq}) is
formally similar to the Schr\"odinger equation for a spin rotating in
magnetic field, the "rate" of back-scattering, which should be
identified with $1/2l$ in our problem, is given by a Fermi golden rule
like expression
\begin{eqnarray}
\frac{1}{2l}= q^2\,{\rm Re}\,\int_{-\infty}^{0} dy \left\langle
e^{2i\epsilon(x-y)-2i\int_{y}^x dzV(z)} \right\rangle_{V(x)}.
\end{eqnarray}
The averaging over the fluctuating potential $V(x)$ with the correlator
(\ref{deltacorrelated-potential}) reduces to the evaluation of a simple
gaussian path integral and indeed leads to the expression (\ref{l}) for
the mean free path.

Note that the preceding derivation requires white-noise-like disorder
with $\sigma_V\xi\ll 1$. We state without proof that, in the limit of
strong disorder $\sigma_V\xi\gtrsim 2$, Eq. (\ref{DMPK}) still holds,
as is made plausible by the consistent derivation of $s=L/l$ in
Appendix \ref{ap:1D}. In this case however, the mean free path has a
different form from Eq. (\ref{l}).

Moreover, the condition (\ref{isot}) implies that the DMPK equation
applies only to modes which lie within a  window $|q|\lesssim
\sqrt{\epsilon^2+\sigma_V^4\xi^2}$. All modes with higher $q$ are
weakly sensitive to disorder and are, in fact, evanescent. When this
window is finite (for finite disorder or finite doping), the evanescent
modes can be ignored provided that the sample is long enough. The precise
condition is $\sigma_V^2L\xi\gg 1$. In this limit, the contribution of
the evanescent modes with $|q|\gtrsim\sqrt{\epsilon^2+\sigma_V^4\xi^2}$
to the conductance is negligible and
 Eq. (\ref{rho1D}) becomes valid.

\section{Analytical results for one-dimensional disorder averaging\label{ap:1D}}

The problem of computing the mean free path $l(q)$ that
enters the 1D distribution [Eq. (\ref{P1D})] is tackled here by
finding analytical solutions for the expectation value of a
certain function of $\lambda$, in our case $\langle \cosh
2\lambda\rangle$, and then adjusting $s=L/l$ in order to
recover that same result from the distribution (\ref{P1D}).
This is done for low dopings and both for the strong and
the weak disorder limits. The procedure is carried out also
for a different function $\langle \cosh^2 2\lambda\rangle$,
which yields an identical mean free path, which confirms
the fact that Eq. (\ref{P1D}) is the exact distribution for
transmissions through graphene with 1D smooth disorder, as modeled by Eq. (\ref{Teq}).

As explained in Ref. \onlinecite{Titov07}, an alternative form of Eq.
(\ref{Teq}) for mode $q$ under 1D disorder is given by Eqs.
(\ref{eq1a}) and (\ref{eq1b}). The dynamics of $\theta$ and $\lambda$
therein is coupled. However, we are interested in expectation values of
functions that do not involve $\theta$ (as is the case of any
observable current cumulant). The following exact reformulation of Eqs.
(\ref{eq1a}) and (\ref{eq1b}) proves useful to obtain them:
\begin{eqnarray}
\vc v(x)&\equiv&\left(\begin{array}{c} \cosh
2\lambda\\\sinh 2\lambda\cos 2\theta\\\sinh 2\lambda\sin
2\theta\end{array}\right),\\
\mat Q(x)&\equiv&\left(\begin{array}{ccc}
0& 2q & 0\\
2q& 0 &-2(\epsilon-V(x))\\
0& 2(\epsilon-V(x))&0
\end{array}\right),\\
\frac{d \vc v(x)}{d x}&=&\mat Q(x)\vc v(x),
\end{eqnarray}
with the exact solution
\begin{equation}
\vc v(L)=P\exp\left(\int_0^L \mat Q(x)dx\right)\vc v(0).
\end{equation}
where $P\exp$ stands for a path ordered exponential. With a
piecewise potential $V(x)$ in which each step of size $\xi$
is statistically independent of the others, the average
$\langle v(L)\rangle$ reads
\begin{equation}
\langle v(L)\rangle=\left[\langle e^{\mat Q
\xi}\rangle\right]^{L/\xi}\vc v(0).\label{avv}
\end{equation}
We therefore need to obtain the average of $\mat U \equiv e^{\mat Q
\xi}$ for a single charge puddle. To proceed analytically we have to
assume that the average doping $\epsilon$ is small, so we expand $\mat
U$ to second order in $\epsilon$. \footnote{Note that this is done only
for analytical convenience. Finite dopings can also be employed by
carrying out this procedure numerically.} We will furthermore assume
that the system contains a large number of puddles, $N=L/\xi\gg 1$. In
this case, as will become apparent later, it is also sufficient to
expand $\mat U$ to second order in $q$, since higher orders will not
contribute to $\langle U\rangle^N$ in the large $N$ limit.

Let us now consider the case of strong disorder, which as
it turns out means $\sigma_V\xi \gtrsim 2$. After the two
previous expansions, $\mat U$ can be averaged by using a
Gaussian distribution for $V$ with dispersion $\sigma_V$.
The resulting expression contains terms of the form
$\mathrm{Erf}(\sqrt{2}\sigma_V\xi)$ and
$e^{-2\sigma_V^2\xi^2}$, which can be greatly simplified to $1$ and $0$ respectively in
the strong disorder limit $\sigma_V\xi\gtrsim 2$. To
exponentiate the result one computes the eigenvalues of the
resulting $\langle \mat U\rangle$. One eigenvalue is zero,
another is small as $q^2$ and the last one is close to 1,
so that if $\mat D$ is the matrix that diagonalizes
$\langle \mat U\rangle$, one can write
\begin{equation}
\left[\langle e^{\mat Q \xi}\rangle\right]^{L/\xi}=\mat D
\left(\begin{array}{ccc}
(1+(q\xi)^2\gamma)^{L/\xi}&0&0\\
0&[(q\xi)^2\tilde\gamma]^{L/\xi}&0\\
0&0&0
\end{array}\right)D^{-1},
\end{equation}
where
\begin{eqnarray}
\gamma&=&\frac{(\sigma_V\xi)^2(2\sqrt{2\pi}\sigma_V\xi+\pi-2)-(\epsilon\xi)^2(\pi+\sqrt{2\pi}\sigma_V\xi)}{2(\sigma_V\xi)^4},\nn\\
\tilde\gamma&=&\frac{\pi(\epsilon\xi)^2-(\sigma_V\xi)^2(\pi-2)}{2(\sigma_V\xi)^4}.\nn
\end{eqnarray}

In the limit of large number of puddles $N=L/\xi\rightarrow
\infty$, we have $[1+(q\xi)^2\gamma]^{L/\xi}\rightarrow
\exp[q^2\xi L \gamma]$, and
$[(q\xi)^2\tilde\gamma]^{L/\xi}\rightarrow 0$, and higher
order corrections in $q$ become irrelevant. By writing down
the expression for $D$ in leading (zeroth) order in $q$,
and selecting the first element of $\langle\vc v(L)\rangle$
in Eq. (\ref{avv}), we arrive at
\begin{equation}
\langle \cosh
2\lambda\rangle=\exp\left(q^2L\xi\gamma\right).
\label{cosh}
\end{equation}
From this result it becomes clear that $q\xi$ is indeed a small expansion parameter for the relevant momenta, since the typical momentum scale that appears after averaging is $q\sim 1/\sqrt{L\xi}\ll 1/\xi$.

Similarly we can obtain $\langle\cosh^22\lambda\rangle$ by doing an
analogous computation for $\vc v_2(L)\equiv\vc v(L)\otimes \vc v(L)$,
whose equation of motion involves $\mat Q\otimes
1\!\!1+1\!\!1\otimes\mat Q$ instead of $\mat Q$. The result in that
case for the first element of $\langle\vc v_2(L)\rangle$ reads
\begin{equation}
\langle \cosh^22\lambda\rangle =
\frac{1}{3}+\frac{2}{3}\exp\left(3 q^2L\xi \gamma\right).
\label{cosh2}
\end{equation}

The above is valid for strong disorder. In the case of weak disorder, we can expand
\begin{eqnarray*}
\langle\mat
e^{\mat Q \xi}\rangle^{L/\xi} &\equiv& \exp\left[\frac{L}{\xi}\log\left(1\!\!1+ \langle Q\rangle \xi+\frac{1}{2}\langle Q^2\rangle\xi^2\right)\right]\\
&\approx&\exp\left\{\frac{L}{\xi}\left[\langle Q\rangle \xi+\frac{1}{2}\xi^2(\langle Q^2\rangle-\langle Q\rangle^2)\right]\right\}.
\end{eqnarray*}
Once again, only one of the eigenvalues of the exponent is relevant for the computation of $\langle \cosh 2\lambda\rangle$. In the limit of large $L/\xi$, this eigenvalue reads $q^2\xi L\gamma$, where now
$$\gamma=\frac{2(\sigma_V\xi)^2}{(\epsilon\xi)^2+(\sigma_V\xi)^4},$$
and we again find
\begin{eqnarray}
\langle \cosh
2\lambda\rangle&=&\exp\left(q^2L\xi\gamma\right),
\nonumber\\
\langle\cosh^22\lambda\rangle&=&\frac{1}{3}+\frac{2}{3}\exp\left(3q^2L\xi\gamma\right).
\label{weak}
\end{eqnarray}

The same averages  computed from the 1D DMPK
distribution [Eq. (\ref{P1D})] read
\begin{eqnarray}
\langle \cosh 2\lambda \rangle &=& \exp(2s)\nonumber\\
\langle \cosh^2 2\lambda \rangle &=& \frac{1}{3}+\frac{2}{3}\exp(6s).
\label{DMPKa}
\end{eqnarray}
%\begin{widetext}
%\begin{eqnarray*}
%\langle \cosh 2\lambda \rangle &=&2\int_0^\infty du\int_0^u
%d\lambda\, \sqrt{\frac{2}{\pi}}\;\frac{e^{-s/4}}{s^{3/2}}
%ue^{-u^2/s} \frac{\sinh 2\lambda\cosh 2\lambda}{\sqrt{\cosh
%2u-\cosh 2\lambda}} \\
%&=&\int_0^\infty du\int_1^{\cosh 2u} dx\,
%\sqrt{\frac{2}{\pi}}\;\frac{e^{-s/4}}{s^{3/2}} ue^{-u^2/s}
%\frac{x}{\sqrt{\cosh 2u-x}}\\
%&=&\frac{2}{\sqrt{\pi}}\;\frac{e^{-s/4}}{s^{3/2}}\int_{-\infty}^\infty
%du\, ue^{-u^2/s} \left( \sinh u\cosh 2u -\frac{2}{3}\sinh^3
%u\right) = \exp(2s)\\
%\langle \cosh^2 2\lambda \rangle &=&2\int_0^\infty
%du\int_0^u d\lambda\,
%\sqrt{\frac{2}{\pi}}\;\frac{e^{-s/4}}{s^{3/2}} ue^{-u^2/s}
%\frac{\sinh 2\lambda\cosh^2 2\lambda}{\sqrt{\cosh 2u-\cosh
%2\lambda}} \\
%&=&\int_0^\infty du\int_1^{\cosh 2u} dx\,
%\sqrt{\frac{2}{\pi}}\;\frac{e^{-s/4}}{s^{3/2}} ue^{-u^2/s}
%\frac{x^2}{\sqrt{\cosh 2u-x}}\\
%&=&\frac{2}{\sqrt{\pi}}\;\frac{e^{-s/4}}{s^{3/2}}\int_{-\infty}^\infty
%du\, ue^{-u^2/s} \left( \sinh u\cosh^2 2u
%-\frac{4}{3}\sinh^3 u\cosh 2u+\frac{4}{5}\sinh^5 u\right) =
%\frac{1}{3}+\frac{2}{3}\exp(6s).
%\end{eqnarray*}
%\end{widetext}
Comparing Eqs. (\ref{cosh})-(\ref{DMPKa}),
we get  $s=\frac{1}{2}q^2L\xi\gamma$, i.e. Eq. (\ref{gamma}).

\bibliography{Puddles}

\begin{thebibliography}{37}
\expandafter\ifx\csname natexlab\endcsname\relax\def\natexlab#1{#1}\fi
\expandafter\ifx\csname bibnamefont\endcsname\relax
  \def\bibnamefont#1{#1}\fi
\expandafter\ifx\csname bibfnamefont\endcsname\relax
  \def\bibfnamefont#1{#1}\fi
\expandafter\ifx\csname citenamefont\endcsname\relax
  \def\citenamefont#1{#1}\fi
\expandafter\ifx\csname url\endcsname\relax
  \def\url#1{\texttt{#1}}\fi
\expandafter\ifx\csname urlprefix\endcsname\relax\def\urlprefix{URL }\fi
\providecommand{\bibinfo}[2]{#2}
\providecommand{\eprint}[2][]{\url{#2}}

\bibitem[{\citenamefont{Novoselov et~al.}(2004)\citenamefont{Novoselov, Geim,
  Morozov, Jiang, Zhang, Dubonos, Grigorieva, and Firsov}}]{Novoselov04}
\bibinfo{author}{\bibfnamefont{K.~S.} \bibnamefont{Novoselov}},
  \bibinfo{author}{\bibfnamefont{A.~K.} \bibnamefont{Geim}},
  \bibinfo{author}{\bibfnamefont{S.~V.} \bibnamefont{Morozov}},
  \bibinfo{author}{\bibfnamefont{D.}~\bibnamefont{Jiang}},
  \bibinfo{author}{\bibfnamefont{Y.}~\bibnamefont{Zhang}},
  \bibinfo{author}{\bibfnamefont{S.~V.} \bibnamefont{Dubonos}},
  \bibinfo{author}{\bibfnamefont{I.~V.} \bibnamefont{Grigorieva}},
  \bibnamefont{and} \bibinfo{author}{\bibfnamefont{A.~A.}
  \bibnamefont{Firsov}}, \bibinfo{journal}{Science}
  \textbf{\bibinfo{volume}{306}}, \bibinfo{pages}{666} (\bibinfo{year}{2004}).

\bibitem[{\citenamefont{Novoselov et~al.}(2005)\citenamefont{Novoselov, Geim,
  Morozov, Jiang, Katsnelson, Grigorieva, Dubonos, and Firsov}}]{Novoselov05}
\bibinfo{author}{\bibfnamefont{K.~S.} \bibnamefont{Novoselov}},
  \bibinfo{author}{\bibfnamefont{A.~K.} \bibnamefont{Geim}},
  \bibinfo{author}{\bibfnamefont{S.~V.} \bibnamefont{Morozov}},
  \bibinfo{author}{\bibfnamefont{D.}~\bibnamefont{Jiang}},
  \bibinfo{author}{\bibfnamefont{M.~I.} \bibnamefont{Katsnelson}},
  \bibinfo{author}{\bibfnamefont{I.~V.} \bibnamefont{Grigorieva}},
  \bibinfo{author}{\bibfnamefont{S.~V.} \bibnamefont{Dubonos}},
  \bibnamefont{and} \bibinfo{author}{\bibfnamefont{A.~A.}
  \bibnamefont{Firsov}}, \bibinfo{journal}{Nature (London)}
  \textbf{\bibinfo{volume}{438}}, \bibinfo{pages}{197} (\bibinfo{year}{2005}).

\bibitem[{\citenamefont{Zhang et~al.}(2005)\citenamefont{Zhang, Tan, Stormer,
  and Kim}}]{Zhang05}
\bibinfo{author}{\bibfnamefont{Y.~B.} \bibnamefont{Zhang}},
  \bibinfo{author}{\bibfnamefont{Y.~W.} \bibnamefont{Tan}},
  \bibinfo{author}{\bibfnamefont{H.~L.} \bibnamefont{Stormer}},
  \bibnamefont{and} \bibinfo{author}{\bibfnamefont{P.}~\bibnamefont{Kim}},
  \bibinfo{journal}{Nature (London)} \textbf{\bibinfo{volume}{438}},
  \bibinfo{pages}{201} (\bibinfo{year}{2005}).

\bibitem[{\citenamefont{Castro~Neto et~al.}(2006)\citenamefont{Castro~Neto,
  Guinea, and Peres}}]{CastroNeto06}
\bibinfo{author}{\bibfnamefont{A.~H.} \bibnamefont{Castro~Neto}},
  \bibinfo{author}{\bibfnamefont{F.}~\bibnamefont{Guinea}}, \bibnamefont{and}
  \bibinfo{author}{\bibfnamefont{N.~M.} \bibnamefont{Peres}},
  \bibinfo{journal}{Physics World} \textbf{\bibinfo{volume}{19}},
  \bibinfo{pages}{33} (\bibinfo{year}{2006}).

\bibitem[{\citenamefont{Katsnelson}(2006)}]{KatsnelsonReview06}
\bibinfo{author}{\bibfnamefont{M.~I.} \bibnamefont{Katsnelson}},
  \bibinfo{journal}{Materials Today} \textbf{\bibinfo{volume}{10}},
  \bibinfo{pages}{20} (\bibinfo{year}{2006}).

\bibitem[{\citenamefont{Geim and Novoselov}(2007)}]{Geim07}
\bibinfo{author}{\bibfnamefont{A.~K.} \bibnamefont{Geim}} \bibnamefont{and}
  \bibinfo{author}{\bibfnamefont{K.~S.} \bibnamefont{Novoselov}},
  \bibinfo{journal}{Nat. Mater.} \textbf{\bibinfo{volume}{6}},
  \bibinfo{pages}{183} (\bibinfo{year}{2007}).

\bibitem[{\citenamefont{Aleiner and Efetov}(2006)}]{Aleiner06}
\bibinfo{author}{\bibfnamefont{I.~L.} \bibnamefont{Aleiner}} \bibnamefont{and}
  \bibinfo{author}{\bibfnamefont{K.~B.} \bibnamefont{Efetov}},
  \bibinfo{journal}{Phys. Rev. Lett.} \textbf{\bibinfo{volume}{97}},
  \bibinfo{pages}{236801} (\bibinfo{year}{2006}).

\bibitem[{\citenamefont{Suzuura and Ando}(2002)}]{Suzuura02}
\bibinfo{author}{\bibfnamefont{H.}~\bibnamefont{Suzuura}} \bibnamefont{and}
  \bibinfo{author}{\bibfnamefont{T.}~\bibnamefont{Ando}},
  \bibinfo{journal}{Phys. Rev. Lett.} \textbf{\bibinfo{volume}{89}},
  \bibinfo{pages}{266603} (\bibinfo{year}{2002}).

\bibitem[{\citenamefont{McCann et~al.}(2006)\citenamefont{McCann, Kechedzhi,
  Fal'ko, Suzuura, Ando, and Altshuler}}]{McCann06}
\bibinfo{author}{\bibfnamefont{E.}~\bibnamefont{McCann}},
  \bibinfo{author}{\bibfnamefont{K.}~\bibnamefont{Kechedzhi}},
  \bibinfo{author}{\bibfnamefont{V.~I.} \bibnamefont{Fal'ko}},
  \bibinfo{author}{\bibfnamefont{H.}~\bibnamefont{Suzuura}},
  \bibinfo{author}{\bibfnamefont{T.}~\bibnamefont{Ando}}, \bibnamefont{and}
  \bibinfo{author}{\bibfnamefont{B.~L.} \bibnamefont{Altshuler}},
  \bibinfo{journal}{Phys. Rev. Lett.} \textbf{\bibinfo{volume}{97}},
  \bibinfo{pages}{146805} (\bibinfo{year}{2006}).

\bibitem[{\citenamefont{Morozov et~al.}(2006)\citenamefont{Morozov, Novoselov,
  Katsnelson, Schedin, Ponomarenko, Jiang, and Geim}}]{Morozov06}
\bibinfo{author}{\bibfnamefont{S.~V.} \bibnamefont{Morozov}},
  \bibinfo{author}{\bibfnamefont{K.~S.} \bibnamefont{Novoselov}},
  \bibinfo{author}{\bibfnamefont{M.~I.} \bibnamefont{Katsnelson}},
  \bibinfo{author}{\bibfnamefont{F.}~\bibnamefont{Schedin}},
  \bibinfo{author}{\bibfnamefont{L.~A.} \bibnamefont{Ponomarenko}},
  \bibinfo{author}{\bibfnamefont{D.}~\bibnamefont{Jiang}}, \bibnamefont{and}
  \bibinfo{author}{\bibfnamefont{A.~K.} \bibnamefont{Geim}},
  \bibinfo{journal}{Phys. Rev. Lett.} \textbf{\bibinfo{volume}{97}},
  \bibinfo{pages}{016801} (\bibinfo{year}{2006}).

\bibitem[{\citenamefont{Morpurgo and Guinea}(2006)}]{Morpurgo06}
\bibinfo{author}{\bibfnamefont{A.~F.} \bibnamefont{Morpurgo}} \bibnamefont{and}
  \bibinfo{author}{\bibfnamefont{F.}~\bibnamefont{Guinea}},
  \bibinfo{journal}{Phys. Rev. Lett.} \textbf{\bibinfo{volume}{97}},
  \bibinfo{pages}{196804} (\bibinfo{year}{2006}).

\bibitem[{\citenamefont{Ostrovsky et~al.}(2006)\citenamefont{Ostrovsky, Gornyi,
  and Mirlin}}]{Ostrovsky06}
\bibinfo{author}{\bibfnamefont{P.~M.} \bibnamefont{Ostrovsky}},
  \bibinfo{author}{\bibfnamefont{I.~V.} \bibnamefont{Gornyi}},
  \bibnamefont{and} \bibinfo{author}{\bibfnamefont{A.~D.}
  \bibnamefont{Mirlin}}, \bibinfo{journal}{Phys. Rev. B}
  \textbf{\bibinfo{volume}{74}}, \bibinfo{pages}{235443}
  (\bibinfo{year}{2006}).

\bibitem[{\citenamefont{Peres et~al.}(2006)\citenamefont{Peres, Guinea, and
  {Castro Neto}}}]{Peres06}
\bibinfo{author}{\bibfnamefont{N.~M.~R.} \bibnamefont{Peres}},
  \bibinfo{author}{\bibfnamefont{F.}~\bibnamefont{Guinea}}, \bibnamefont{and}
  \bibinfo{author}{\bibfnamefont{A.~H.} \bibnamefont{{Castro Neto}}},
  \bibinfo{journal}{Phys. Rev. B} \textbf{\bibinfo{volume}{73}},
  \bibinfo{pages}{125411} (\bibinfo{year}{2006}).

\bibitem[{\citenamefont{Ziegler}(2006)}]{Ziegler06}
\bibinfo{author}{\bibfnamefont{K.}~\bibnamefont{Ziegler}},
  \bibinfo{journal}{Phys. Rev. Lett.} \textbf{\bibinfo{volume}{97}},
  \bibinfo{pages}{266802} (\bibinfo{year}{2006}).

\bibitem[{\citenamefont{Nomura and MacDonald}(2007)}]{Nomura07}
\bibinfo{author}{\bibfnamefont{K.}~\bibnamefont{Nomura}} \bibnamefont{and}
  \bibinfo{author}{\bibfnamefont{A.~H.} \bibnamefont{MacDonald}},
  \bibinfo{journal}{Phys. Rev. Lett.} \textbf{\bibinfo{volume}{98}},
  \bibinfo{pages}{076602} (\bibinfo{year}{2007}).

\bibitem[{\citenamefont{Meyer et~al.}(2007)\citenamefont{Meyer, Geim,
  Katsnelson, Novoselov, Booth, and Roth}}]{Meyer07}
\bibinfo{author}{\bibfnamefont{J.~C.} \bibnamefont{Meyer}},
  \bibinfo{author}{\bibfnamefont{A.~K.} \bibnamefont{Geim}},
  \bibinfo{author}{\bibfnamefont{M.~I.} \bibnamefont{Katsnelson}},
  \bibinfo{author}{\bibfnamefont{K.~S.} \bibnamefont{Novoselov}},
  \bibinfo{author}{\bibfnamefont{T.~J.} \bibnamefont{Booth}}, \bibnamefont{and}
  \bibinfo{author}{\bibfnamefont{S.}~\bibnamefont{Roth}},
  \bibinfo{journal}{Nature (London)} \textbf{\bibinfo{volume}{446}},
  \bibinfo{pages}{60} (\bibinfo{year}{2007}).

\bibitem[{\citenamefont{Galitsky et~al.}(2007)\citenamefont{Galitsky,
  Shaffique, and Das~Sarma}}]{Galitski07}
\bibinfo{author}{\bibfnamefont{V.~M.} \bibnamefont{Galitsky}},
  \bibinfo{author}{\bibfnamefont{A.}~\bibnamefont{Shaffique}},
  \bibnamefont{and} \bibinfo{author}{\bibfnamefont{S.}~\bibnamefont{Das~Sarma}}
  (\bibinfo{year}{2007}), \bibinfo{note}{arXiv:cond-mat/0702117, Phys. Rev. B
  (to be published)}.

\bibitem[{\citenamefont{Hwang et~al.}(2007)\citenamefont{Hwang, Adam, and
  Das~Sarma}}]{Hwang07}
\bibinfo{author}{\bibfnamefont{E.~H.} \bibnamefont{Hwang}},
  \bibinfo{author}{\bibfnamefont{S.}~\bibnamefont{Adam}}, \bibnamefont{and}
  \bibinfo{author}{\bibfnamefont{S.}~\bibnamefont{Das~Sarma}},
  \bibinfo{journal}{Phys. Rev. Lett.} \textbf{\bibinfo{volume}{98}},
  \bibinfo{pages}{186806} (\bibinfo{year}{2007}).

\bibitem[{\citenamefont{Cho and Fuhrer}(2007)}]{Cho07}
\bibinfo{author}{\bibfnamefont{S.}~\bibnamefont{Cho}} \bibnamefont{and}
  \bibinfo{author}{\bibfnamefont{M.~S.} \bibnamefont{Fuhrer}}
  (\bibinfo{year}{2007}), \bibinfo{note}{arXiv:0705.3239}.

\bibitem[{\citenamefont{Katsnelson et~al.}(2006)\citenamefont{Katsnelson,
  Novoselov, and Geim}}]{KatsnelsonKlein06}
\bibinfo{author}{\bibfnamefont{M.~I.} \bibnamefont{Katsnelson}},
  \bibinfo{author}{\bibfnamefont{K.~S.} \bibnamefont{Novoselov}},
  \bibnamefont{and} \bibinfo{author}{\bibfnamefont{A.~K.} \bibnamefont{Geim}},
  \bibinfo{journal}{Nat. Phys.} \textbf{\bibinfo{volume}{2}},
  \bibinfo{pages}{620} (\bibinfo{year}{2006}).

\bibitem[{\citenamefont{Martin et~al.}(2007)\citenamefont{Martin, Akerman,
  Ulbricht, Lohmann, Smet, von Klitzing, and Yacoby}}]{Martin07}
\bibinfo{author}{\bibfnamefont{J.}~\bibnamefont{Martin}},
  \bibinfo{author}{\bibfnamefont{N.}~\bibnamefont{Akerman}},
  \bibinfo{author}{\bibfnamefont{G.}~\bibnamefont{Ulbricht}},
  \bibinfo{author}{\bibfnamefont{T.}~\bibnamefont{Lohmann}},
  \bibinfo{author}{\bibfnamefont{J.~H.} \bibnamefont{Smet}},
  \bibinfo{author}{\bibfnamefont{K.}~\bibnamefont{von Klitzing}},
  \bibnamefont{and} \bibinfo{author}{\bibfnamefont{A.}~\bibnamefont{Yacoby}}
  (\bibinfo{year}{2007}), \bibinfo{note}{arXiv:0705.2180v1}.

\bibitem[{\citenamefont{Castro~Neto and Kim}(2007)}]{CastroNeto07}
\bibinfo{author}{\bibfnamefont{A.~H.} \bibnamefont{Castro~Neto}}
  \bibnamefont{and} \bibinfo{author}{\bibfnamefont{E.~A.} \bibnamefont{Kim}}
  (\bibinfo{year}{2007}), \bibinfo{note}{arXiv:cond-mat/0702562}.

\bibitem[{\citenamefont{Ostrovsky et~al.}(2007)\citenamefont{Ostrovsky, Gornyi,
  and Mirlin}}]{Ostrovsky07}
\bibinfo{author}{\bibfnamefont{P.~M.} \bibnamefont{Ostrovsky}},
  \bibinfo{author}{\bibfnamefont{I.~V.} \bibnamefont{Gornyi}},
  \bibnamefont{and} \bibinfo{author}{\bibfnamefont{A.~D.}
  \bibnamefont{Mirlin}}, \bibinfo{journal}{Phys. Rev. Lett.}
  \textbf{\bibinfo{volume}{98}}, \bibinfo{pages}{256801}
  (\bibinfo{year}{2007}).

\bibitem[{\citenamefont{Bardarson et~al.}(2007)\citenamefont{Bardarson,
  Tworzydlo, Brouwer, and Beenakker}}]{Bardarson07}
\bibinfo{author}{\bibfnamefont{J.~H.} \bibnamefont{Bardarson}},
  \bibinfo{author}{\bibfnamefont{J.}~\bibnamefont{Tworzydlo}},
  \bibinfo{author}{\bibfnamefont{P.~W.} \bibnamefont{Brouwer}},
  \bibnamefont{and} \bibinfo{author}{\bibfnamefont{C.~W.~J.}
  \bibnamefont{Beenakker}}, \bibinfo{journal}{Phys. Rev. Lett.}
  \textbf{\bibinfo{volume}{99}}, \bibinfo{pages}{106801}
  (\bibinfo{year}{2007}).

\bibitem[{\citenamefont{Nomura et~al.}(2007)\citenamefont{Nomura, Koshino, and
  Ryu}}]{NomuraBeta07}
\bibinfo{author}{\bibfnamefont{K.}~\bibnamefont{Nomura}},
  \bibinfo{author}{\bibfnamefont{M.}~\bibnamefont{Koshino}}, \bibnamefont{and}
  \bibinfo{author}{\bibfnamefont{S.}~\bibnamefont{Ryu}} (\bibinfo{year}{2007}),
  \bibinfo{note}{arXiv:0705.1607v1}.

\bibitem[{\citenamefont{Titov}(2007)}]{Titov07}
\bibinfo{author}{\bibfnamefont{M.}~\bibnamefont{Titov}},
  \bibinfo{journal}{Europhys. Lett.} \textbf{\bibinfo{volume}{79}},
  \bibinfo{pages}{17004} (\bibinfo{year}{2007}).

\bibitem[{\citenamefont{Rycerz et~al.}(2007)\citenamefont{Rycerz, Tworzydlo,
  and Beenakker}}]{Rycerz07}
\bibinfo{author}{\bibfnamefont{A.}~\bibnamefont{Rycerz}},
  \bibinfo{author}{\bibfnamefont{J.}~\bibnamefont{Tworzydlo}},
  \bibnamefont{and} \bibinfo{author}{\bibfnamefont{C.~W.~J.}
  \bibnamefont{Beenakker}}, \bibinfo{journal}{Europhys. Lett.}
  \textbf{\bibinfo{volume}{79}}, \bibinfo{pages}{57003} (\bibinfo{year}{2007}).

\bibitem[{\citenamefont{Cheianov et~al.}(2007)\citenamefont{Cheianov, Fal'ko,
  Altshuler, and Aleiner}}]{Cheianov07}
\bibinfo{author}{\bibfnamefont{V.~V.} \bibnamefont{Cheianov}},
  \bibinfo{author}{\bibfnamefont{V.~I.} \bibnamefont{Fal'ko}},
  \bibinfo{author}{\bibfnamefont{B.~L.} \bibnamefont{Altshuler}},
  \bibnamefont{and} \bibinfo{author}{\bibfnamefont{I.~L.}
  \bibnamefont{Aleiner}} (\bibinfo{year}{2007}),
  \bibinfo{note}{arXiv:0706.2968v2}.

\bibitem[{\citenamefont{Beenakker}(1997)}]{BeenakkerRMP97}
\bibinfo{author}{\bibfnamefont{C.~W.~J.} \bibnamefont{Beenakker}},
  \bibinfo{journal}{Rev. Mod. Phys.} \textbf{\bibinfo{volume}{69}},
  \bibinfo{pages}{731} (\bibinfo{year}{1997}).

\bibitem[{\citenamefont{Ryu et~al.}(2007)\citenamefont{Ryu, Mudry, Obuse, and
  Furusaki}}]{Ryu07}
\bibinfo{author}{\bibfnamefont{S.}~\bibnamefont{Ryu}},
  \bibinfo{author}{\bibfnamefont{C.}~\bibnamefont{Mudry}},
  \bibinfo{author}{\bibfnamefont{H.}~\bibnamefont{Obuse}}, \bibnamefont{and}
  \bibinfo{author}{\bibfnamefont{A.}~\bibnamefont{Furusaki}},
  \bibinfo{journal}{Phys. Rev. Lett.} \textbf{\bibinfo{volume}{99}},
  \bibinfo{pages}{116601} (\bibinfo{year}{2007}).

\bibitem[{\citenamefont{Tan et~al.}(2007)\citenamefont{Tan, Zhang, Bolotin,
  Zhao, Adam, Hwang, Sarma, Stormer, and Kim}}]{Tan07}
\bibinfo{author}{\bibfnamefont{Y.-W.} \bibnamefont{Tan}},
  \bibinfo{author}{\bibfnamefont{Y.}~\bibnamefont{Zhang}},
  \bibinfo{author}{\bibfnamefont{K.}~\bibnamefont{Bolotin}},
  \bibinfo{author}{\bibfnamefont{Y.}~\bibnamefont{Zhao}},
  \bibinfo{author}{\bibfnamefont{S.}~\bibnamefont{Adam}},
  \bibinfo{author}{\bibfnamefont{E.}~\bibnamefont{Hwang}},
  \bibinfo{author}{\bibfnamefont{S.~D.} \bibnamefont{Sarma}},
  \bibinfo{author}{\bibfnamefont{H.~L.} \bibnamefont{Stormer}},
  \bibnamefont{and} \bibinfo{author}{\bibfnamefont{P.}~\bibnamefont{Kim}}
  (\bibinfo{year}{2007}), \bibinfo{note}{arXiv:0707.1807v1}.

\bibitem[{\citenamefont{Mello et~al.}(1988)\citenamefont{Mello, Pereyra, and
  Kumar}}]{Mello88}
\bibinfo{author}{\bibfnamefont{P.~A.} \bibnamefont{Mello}},
  \bibinfo{author}{\bibfnamefont{P.}~\bibnamefont{Pereyra}}, \bibnamefont{and}
  \bibinfo{author}{\bibfnamefont{N.}~\bibnamefont{Kumar}},
  \bibinfo{journal}{Annals of Physics} \textbf{\bibinfo{volume}{181}},
  \bibinfo{pages}{290} (\bibinfo{year}{1988}).

\bibitem[{\citenamefont{Cheianov and Fal'ko}(2006)}]{Cheianov06pn}
\bibinfo{author}{\bibfnamefont{V.~V.} \bibnamefont{Cheianov}} \bibnamefont{and}
  \bibinfo{author}{\bibfnamefont{V.~I.} \bibnamefont{Fal'ko}},
  \bibinfo{journal}{Phys. Rev. B} \textbf{\bibinfo{volume}{74}},
  \bibinfo{pages}{041403(R)} (\bibinfo{year}{2006}).

\bibitem[{\citenamefont{Abrikosov}(1981)}]{Abrikosov81}
\bibinfo{author}{\bibfnamefont{A.~A.} \bibnamefont{Abrikosov}},
  \bibinfo{journal}{Solid State Communications} \textbf{\bibinfo{volume}{37}},
  \bibinfo{pages}{997} (\bibinfo{year}{1981}).

\bibitem[{\citenamefont{Tworzydlo et~al.}(2006)\citenamefont{Tworzydlo,
  Trauzettel, Titov, Rycerz, and Beenakker}}]{Tworzydlo06}
\bibinfo{author}{\bibfnamefont{J.}~\bibnamefont{Tworzydlo}},
  \bibinfo{author}{\bibfnamefont{B.}~\bibnamefont{Trauzettel}},
  \bibinfo{author}{\bibfnamefont{M.}~\bibnamefont{Titov}},
  \bibinfo{author}{\bibfnamefont{A.}~\bibnamefont{Rycerz}}, \bibnamefont{and}
  \bibinfo{author}{\bibfnamefont{C.~W.~J.} \bibnamefont{Beenakker}},
  \bibinfo{journal}{Phys. Rev. Lett.} \textbf{\bibinfo{volume}{96}},
  \bibinfo{pages}{246802} (\bibinfo{year}{2006}).

\bibitem[{\citenamefont{Dorokhov}(1984)}]{Dorokhov84}
\bibinfo{author}{\bibfnamefont{O.~N.} \bibnamefont{Dorokhov}},
  \bibinfo{journal}{Solid State Communications} \textbf{\bibinfo{volume}{51}},
  \bibinfo{pages}{381} (\bibinfo{year}{1984}).

\bibitem[{\citenamefont{Nazarov}(1994)}]{Nazarov94}
\bibinfo{author}{\bibfnamefont{Y.~V.} \bibnamefont{Nazarov}},
  \bibinfo{journal}{Phys. Rev. Lett.} \textbf{\bibinfo{volume}{73}},
  \bibinfo{pages}{134} (\bibinfo{year}{1994}).

\end{thebibliography}

\end{document}